\documentstyle[psfig,aaspp4,12pt,multicol]{article}

\lefthead{J.~M.~Miller et al.}  \righthead{Cygnus X-1}

\received{Date}
\revised{Date}
\accepted{Date}

\journalid{vol}{date}
\articleid{1}{4}
\paperid{id}

\cpright{AAS}{1999}
\ccc{x}

\begin{document}

\title{Resolving the Composite Fe~K$\alpha$ Emission Line in the
Galactic Black Hole Cygnus X-1 with Chandra}

\author{J.~M.~Miller\altaffilmark{1},
        A.~C.~Fabian\altaffilmark{2},
	R.~Wijnands\altaffilmark{1,4},
	R.~A.~Remillard\altaffilmark{1},
	P.~Wojdowski\altaffilmark{1},
	N.~S.~Schulz\altaffilmark{1},
	T.~Di~Matteo\altaffilmark{3},
	H.~L.~Marshall\altaffilmark{1},
	C.~R.~Canizares\altaffilmark{1},
	D.~Pooley\altaffilmark{1},
	W.~H.~G.~Lewin\altaffilmark{1}
	}

\altaffiltext{1}{Center~for~Space~Research and Department~of~Physics,
        Massachusetts~Institute~of~Technology, Cambridge, MA
        02139--4307; jmm@space.mit.edu}
\altaffiltext{2}{Institute of Astronomy, University of Cambridge,
        Madingley Road, Cambridge CB3 OHA, UK}
\altaffiltext{3}{Max-Planck-Institut fur Astrophysik,
        Karl-Schwarzschild-Str.~1, Postfach 1317, D85741 Garching, DE}
\altaffiltext{4}{\it Chandra Fellow}

\keywords{Black hole physics -- relativity -- stars: binaries
(Cygnus~X-1) -- physical data and processes: accretion disks --
X-rays: stars}

\authoremail{jmm@space.mit.edu}

\label{firstpage}

\begin{abstract}
We observed the Galactic black hole Cygnus X-1 with the
\textit{Chandra} High Energy Transmission Grating Spectrometer for 30
kiloseconds on 2001 January 4.  The source was in an intermediate
state, with a flux that was approximately twice that commonly observed
in its persistent low/hard state.  Our best-fit model for the X-ray
spectrum includes narrow Gaussian emission line (E$=6.415\pm
0.007$~keV, FWHM$=80^{+28}_{-19}$~eV, W$=16^{+3}_{-2}$~eV) and broad
line (E=$5.82^{+0.06}_{-0.07}$~keV, FWHM$=1.9^{+0.5}_{-0.3}$~keV,
W$=140^{+70}_{-40}$~eV) components, and a smeared edge at $7.3\pm
0.2$~keV ($\tau \sim 1.0$).  The broad line profile is not as strongly
skewed as those observed in some Seyfert galaxies.  We interpret these
features in terms of an accretion disk with irradiation of the inner
disk producing a broad Fe~K$\alpha$ emission line and edge, and
irradiation of the outer disk producing a narrow Fe~K$\alpha$ emission
line.  The broad line is likely shaped predominantly by Doppler shifts
and gravitational effects, and to a lesser degree by Compton
scattering due to reflection.  We discuss the underlying continuum
X-ray spectrum and these line features in the context of diagnosing
the accretion flow geometry in Cygnus X-1 and other Galactic black
holes.
\end{abstract}


\section{Introduction}
Cygnus X-1 was the first object to be classified as an X-ray binary
system, and also the first to be recognized as a black hole, via
optical radial velocity measurements (Webster \& Murdin, 1971; Bolton,
1972).  Correlated X-ray and radio intensity variations led to the
identification of ``X-ray states'' --- periods of high X-ray intensity
and spectral softness, or low X-ray intensity and spectral hardness
--- in Cygnus X-1 (Tananbaum et al. 1972).  Since then, states have
become an essential part of how Galactic black holes are discussed and
understood (for reviews, see Tanaka \& Lewin 1995 and Done 2001; see
also Homan et al. 2001).  Cygnus X-1 has been active in X-rays since
its discovery, and presently is the only known persistent Galactic
black hole system with a high mass companion (an O9.7 Iab supergiant;
Gies et al. 1982, 1986).  The black hole mass is dynamically
constrained to a lower limit of $M_{BH}\geq9.5~M_{\odot}$ (Paczy\'nski
1974); Herrero et al. (1995) suggest that the most probable mass is
$M_{BH}\geq10.1~M_{\odot}$.

The bright, persistent nature of Cygnus X-1, its constrained
inclination ($i \simeq 35^{\circ}$, Gies \& Bolton 1986), modest
distance (2.5~kpc, Bregman et al. 1973), and state transitions have
made it a favorite source for observers and theorists.  In particular,
it is a testbed for models aiming to describe links between states and
accretion flow geometries, and how these are related to the mass
accretion rate ($\dot{m}$).  In general, ``high/soft'' states are
associated with high $\dot{m}$, and ``low/hard'' states are associated
with low $\dot{m}$.  State identifications usually also rely upon fast
X-ray variability analysis.

A variety of models connecting states and the accretion geometry have
performed well, but two show particular promise while making very
different predictions.  Esin et al. (1998) have described the behavior
of Cygnus X-1 in terms of an advection-dominated accretion flow
(ADAF).  This model predicts that when $\dot{m}$ is relatively high,
the disk may extend to the marginally stable orbit (the high/soft
state).  When $\dot{m}$ is lower, the disk is recessed, and the inner
region is a hot, quasi-spherical, optically-thin ADAF.  In contrast,
Young et al. (2001) have described the behavior of Cygnus X-1 in terms
of a disk which always extends to the marginally stable orbit but with a
changing surface ionization that determines the X-ray state.

Fe~K$\alpha$ emission lines may trace bulk velocities, temperatures,
and even strong gravitational effects.  Therefore, such lines may
serve as tools for placing observational constraints on accretion flow
models.  In some Seyfert galaxies, characteristic broad (sometimes
double-peaked) Fe~K$\alpha$ emission lines have demonstrated the
presence of an accretion disk down to the innermost stable circular
orbit around the central black hole (see, e.g. Weaver, Gelbord, \&
Yaqoob 2001).  In Galactic black holes, line profiles are often less
distinct; among these systems, the lines observed in Cygnus X-1 are
among the strongest.  Barr, White, and Page (1985) discovered the
broad Fe~K$\alpha$ line (E$=$6.2~keV, FWHM$=$1.2~keV, W$=$120~eV) in
an \textit{EXOSAT} spectrum of Cygnus X-1.  This measured centroid
energy was slightly lower than the neutral value of 6.40~keV (Kaastra
\& Mewe 1993).  Because such features are broad and relatively weak
compared to lines in Seyfert galaxies, one might worry that the
profiles are artifacts of poor continuum models or poor instrumental
response.  However, an Fe~K$\alpha$ line has been required to obtain
statistically acceptable fits to spectra observed from Cygnus X-1 with
a number of instruments, and for a variety of continuum models and
source luminosities (for recent \textit{ASCA} results, see Ebisawa et
al. 1996 and Cui et al. 1998; for \textit{BeppoSAX} results, see
Di~Salvo et al. 2001 and Frontera et al. 2001).

In 2000 November, Cygnus X-1 entered a high-intensity, spectrally-soft
state (see, e. g., Cui, Feng, \& Ertmer 2002; Pottschmidt et
al. 2002).  Although short high/soft states are often seen in this
source, extended high/soft states are rare, previously occurring in
1980 (Ogawara et al. 1982) and 1996 (Cui 1996).  The \textit{Chandra}
High Energy Transmission Grating Spectrometer (HETGS; Canizares et
al. 2002, in prep.) is uniquely suited to resolving structure within
broad Fe~K$\alpha$ lines.  To test models for how accretion geometries
change with source intensity (and therefore $\dot{m}$), we requested a
\textit{Chandra} observation of Cygnus X-1 to obtain a high-resolution
spectrum in the Fe~K$\alpha$ line region during this high intensity
state.  We were granted a Director's Discretionary Time observation
for this purpose and present results from that observation here.

Section 2 describes the instrumental configuration we used to observe
Cygnus X-1.  Section 3 details the models we used to fit the spectrum;
the fit results are presented in Section 4.  We discuss the
implications of our fits for accretion flow models and compare our
results on Cygnus X-1 to other Galactic black holes in Section 4.
Finally, we summarize the main points of our work in Section 5.
Future work will focus on the rich line spectrum below 2~keV,
which has been noted in related work (Miller et al. 2001a, 2001b; see
also Schulz et al. 2001, Marshall et al. 2001a), and modeling the
broad-band X-ray spectrum with a variety of reflection models via
simultaneous \textit{RXTE} observations.  Preliminary results from
fits with the ``constant-density ionized disk model'' (Ross, Fabian,
\& Young 1999; hereafter RFY) to the 0.65-100~keV spectrum indicate an
ionized accretion disk: log$(\xi)=3.0-3.5$~erg~cm~s$^{-1}$, (where
$\xi=L_{X}/nR^{2}$, $\xi$ is the ionization parameter, $L_{X}$ is the
X-ray luminosity, $n$ is the hydrogen number density, and $R$ is
radius).  We also find that 40-50\% of the observed X-ray flux may be
reflected from the accretion disk.

\section{Observation and Instrumental Configuration}
We observed Cygnus X-1 on 2001 January 4, from 06:03:47 to 14:49:20
(UT) -- a total of 32.1 ks.  Near the time of our observation, the
source was flaring to some of the highest intensity levels since the
1996 ``high/soft'' state (see Figure 1).  We identify the source state
we observed as an ``intermediate'' state (see Section 4.1).  The
observatory was still slewing onto the source during the first 0.9 ks
of this observation; this time span in not included for spectral
analysis.  A 0.5~ks dip in the X-ray lightcurve occurs 25.7~ks into
this observation; during this time the count-rate drops to
approximately half of the mean rate (such dips are common in Cygnus
X-1; see Ba\l{}uci\'nska-Church et al. 2000).  We do not include data
from within the dip for this analysis; the characteristics of the dip
will be reported in a separate paper.  The remaining 30.7~ks was
selected for analysis.  Based on the ephemeris of La Sala et
al. (1998), this observation spanned a binary phase of
$\phi=0.76-0.82$ in the 5.6-day orbital period (with the black hole
moving away from us; for a second recent ephemeris, see Brocksopp et
al. 1999).

We used the HETGS to disperse the incident X-ray flux, which was
read-out by the Advanced CCD Imaging Spectrometer (ACIS) spectroscopic
array (ACIS-S) in continuous-clocking (CC) mode.  This step was taken
to reduce photon pile-up in the dispersed spectrum: the frametime in
CC mode is about 2.8~ms, far less than the nominal 3.2 s in standard
modes.  Approximately half of the incident photons are not dispersed
by the HETGS.  We prevented these zeroth-order photons from being
read-out with a 100-column blocking filter on the ACIS-S3 CCD.  This
step was taken to avoid possible telemetry saturation and dropped CCD
frames.  The aimpoint was moved +4 mm in the Z-direction (away from
the read-out nodes) to prevent possible radiation damage to the
nominal aimpoint.  A Y-coordinate translation of $-80$ arcseconds was
used to place as much of the dispersed spectrum on the ACIS-S3 CCD as
possible.

At the time we conducted our analysis, the standard CIAO processing
tools were unable to handle data taken in this mode.  We have
developed a robust set of custom processing routines.  This processing
method is described in detail in a paper reporting results from a
\textit{Chandra} HETGS observation of the Rapid Burster in outburst
(Marshall et al. 2001b).  All aspects of how the spectral grating
orders and background were selected and how the data were corrected
for the instrument response are as those previously reported.  The
only important difference between that observation and our observation
of Cygnus X-1 is that the zeroth-order was blocked.  The location of
the zeroth order is needed to determine event wavelengths, which are a
function of the dispersion distance.  Without a precise measure of the
location of the zeroth order, we instead use a different method to
establish this relationship.  We examined the locations of the neutral
Si absorption edges (due to the Si-based CCDs) in opposite grating
orders.  In raw counts spectra, we were able to determine the zeroth
order position by iterating its location until the Si edges appeared
at the same wavelength in every order.  This position was fine-tuned
by fitting the most prominent absorption lines in opposite grating
orders and requiring that the centroid wavelengths agree to within
0.05\% uncertainty (0.05\% is the uncertainty in the official HETGS
wavelength calibration; Canizares et al. 2002, in prep.).  We are
confident that our wavelength calibration is equivalent to that for
standard observing modes.

We estimate that photon pile-up in our observation is negligible.
Marshall et al. (2001a) report that systematic flux errors across the
HETGS band are likely less than 5\% in this mode.  Future observations
in this mode will enable better calibration and will likely make
significant improvements.  We find a ``jump'' at 2.05~keV previously
seen in the spectra of bright sources observed with \textit{Chandra}
(e.g. Miller et al. 2001c; Patel et al. 2001, Juett et al. 2002).  We
regard this as an instrumental effect and fit it with an inverse edge
(``$\tau$''$=-0.176$).  We also make note of two single-bin features
near 1.85~keV (6.7 $\AA$), which are also instrumental artifacts.

\section{Analysis}
We considered the four first-order dispersed spectra for this
analysis: the two High Energy Grating (HEG) and the two Medium Energy
Grating (MEG) spectra on opposite sides of the zeroth order
(hereafter, the $+1$ and $-1$ orders).  An examination of these
spectra revealed that the HEG~$+$1 and MEG~$-$1 orders are less
affected by the spectrum dithering off the CCD array and relative
CCD gain differences than their counterparts (due to the Z-coordinate
translation).  In characterizing the broad band spectrum, we therefore
considered only these grating orders, with 5\% systematic flux errors
added in quadrature.  We exclude data affected by chip gaps.  Within
these narrow energy ranges we included data from the HEG~$-$1 and
MEG~$+$1 orders, normalized to the continuum level in the HEG~$+$1 and
MEG~$-$1 orders.

Although we examined the spectrum at energies as low as 0.5~keV to
understand how our absorption model fit the photoelectric oxygen
absorption edge from the inter-stellar medium (ISM, see below), the
MEG~$-1$ spectrum is poorly detected below 0.65~keV.  Therefore, we
only fit the MEG~$-1$ spectrum at energies above 0.65~keV.  For the
same reason, we only considered energies above 1.0~keV in fitting the
HEG~$+1$ spectrum.  At 2.4~keV, the MEG~$-1$ order has a chip gap, and
above this energy the HEG has a higher effective area; we only
consider the MEG spectra for energies below 2.4~keV.  The effective
upper-limit to the sensitivity of the HEG is 10.0~keV; our fits to the
HEG~$+1$ extended to this energy.

The spectra were fit using XSPEC version 11.1.0 (Arnaud 1996).  The
MEG~$-1$ and HEG~$+1$ spectra were fit jointly, allowing an overall
normalization constant to float between them (generally, the constant
indicated that the normalization agreed to within 2\% or better).
Systematic errors were added to the flux values using the FTOOL
``grppha.''  Significances quoted in this paper were calculated using
the F-statistic with the ``ftest'' task within XSPEC.

All spectral models were multiplied by a model for photoelectric
absorption in the ISM, with variable elemental abundances.  Local fits
to the absorption edges agree with the absorption model reported in a
previous \textit{Chandra} observation of Cygnus~X-1 by Schulz et
al. (2001).  We used the ``vphabs'' model in all fits with the
abundances adjusted to agree with Schulz et al. (2001), with the minor
distinction that we used the Verner et al. (1993) cross-sections to
fit the Fe~L3 edge, and that we found no evidence for the Fe~L1 edge.
Several components of this model should be noted: the neutral hydrogen
column density is $6.21\times10^{21}~{\rm cm}^{-2}$, oxygen is 7\%
under-abundant (relative to solar) and a better fit is obtained for an
edge at 0.536~keV rather than the predicted 0.532~keV, iron is 25\%
under-abundant, neon is 11\% over-abundant, and all other edges are
consistent with solar values (relative to the abundances stated by
Morrison \& McCammon 1983; $A_{Fe}/A_{H} \simeq 3.3 \times 10^{-5}$)
and expected wavelengths.  These absorption values were not allowed to
vary in fits to the 0.65--10.0~keV spectrum.

A wide variety of models have been used to fit the X-ray spectrum of
Cygnus X-1.  These range from models which attempt to describe
Compton-upscattering of seed photons in a corona, to more
phenomenological models.  Often, these cannot be distinguished on the
basis of a goodness-of-fit statistic (see, e.g., Nowak, Wilms, \& Dove
2002).  As an example of the former, we attempted to fit the observed
0.65-10.0~keV spectrum with the ``compTT'' model (Titarchuk et
al. 1994).  We also fit the spectrum with a model consisting of the
multicolor disk blackbody (hereafter, MCD; Mitsuda et al. 1984) model
and a power-law.  This additive model is commonly a good fit to
Galactic black hole spectra, and provides a standard for comparison to
other sources.  The breadth and strength of Fe~K$\alpha$ line and K
edge features are sufficient to affect fits to the underlying
continuum, and so we have analyzed the continuum and Fe~K$\alpha$ line
region jointly.

\section{Results}
\subsection{The Continuum Spectrum}
In Figure 2, we show the results of fitting an MCD plus power-law
model to the spectra (Model 1 in Table 1, but without a narrow
Gaussian line component).  A soft, thermal component is not required
in some low/hard state spectra observed with other instruments.
However, in this observation such a component is strongly required.
In Panels A and B of Figure 2, structure is apparent in the
Fe~K$\alpha$ line region -- most notably a very narrow emission line.

Fits with the compTT model were not statistically acceptable.  The
seed photon temperature and electron temperature of the
Compton-upscattering corona could not be constrained (errors on these
parameters were several times larger than the values measured).
Fitting only this model, $\chi^{2}/\nu > 8$ (where $\nu$ is the number
of degrees of freedom in the fit).  Fitting an MCD component
simultaneously with compTT yielded a better but still unacceptable fit
($\chi^{2}/\nu > 3$).  Moreover, the data/model ratio shows the same
structure in the Fe~K$\alpha$ line region that is seen in Panel B of
Figure 2.  Hereafter, we restrict our discussion to MCD plus power-law
models.  We note that compTT may still be a good description of
spectra from Cygnus X-1 in different states.

In Table 1, we list the parameters obtained by fitting a series of MCD
plus power-law models with different local models for the
Fe~K$\alpha$ line region.  Model 1 includes a narrow Gaussian to
fit the narrow line evident in Panel B of Figure 2 (see also Figures 3
and 4).  Model 2 adds a smeared edge component (``smedge,'' Ebisawa et
al. 1994).  Model 3 adds a Gaussian to Model 2 to fit a broad emission
line.  Model 4 adds the ``diskline'' model (Fabian et al. 1989) to
Model 2 instead of a Gaussian.  The diskline model explicitly takes
into account the Doppler and general relativistic shifts expected for
a line produced via the irradiation of an accretion disk near to a
black hole.

These models yield apparently poor fits: $\chi^{2}/\nu \simeq$ 1.8.
This is due to the fact that we have not fit a model for the complex
absorption spectrum below 2~keV (Miller et al. 2001a, 2001b; Schulz et
al. 2001; Marshall et al. 2001b).  We include this energy range
because it is critical for accurately characterizing the overall
continuum shape.  Fitting the same models to the spectrum in the
1.8--10.0~keV band (chosen to include the instrumental jump at
2.05~keV), statistically acceptable fits are obtained: $\chi^{2}/\nu
\simeq$0.85--1.15.  Thus, we believe that the fits obtained on the
0.65--10.0~keV band are meaningful.

The peak color temperatures of the inner accretion disk measured via
the MCD model ($kT=0.236\pm 0.002$ keV with Model 3) are above those
which have been measured in the low/hard state (e.g. Ebisawa et
al. 1996), and below those measured in the high/soft state (e.g. Cui
et al. 1998).  The power-law indices we measure ($\Gamma = 1.84\pm
0.01$ with Model 3) are similarly intermediate.  Assuming a distance
of 2.5~kpc to Cygnus X-1, we measure $L_{X}=1.03\pm 0.02 \times
10^{37}$~ergs/s (0.5-10.0 keV) with Model 3.  This luminosity is
approximately twice that commonly measured in low/hard state, but not
as high as the luminosities observed during the high/soft states
observed previously in 1980 and 1996.  Indeed, Belloni et al. (1996)
identify the activity in 1996 as an ``intermediate'' state based
partially on a blackbody temperature of $kT=0.36\pm 0.01$~keV and a
power-law index of $\Gamma=2.15\pm 0.02$ (the latter being
intermediate between canonical low/hard and high/soft state values).
As the spectrum and luminosity we have observed with \textit{Chandra}
are between typical low/hard and high/soft states values, we
characterize this as an ``intermediate'' state.  However we note that
it is a different kind of intermediate state than described by Belloni
et al. (1996) in that it is spectrally harder.

If the distance towards a source and its inclination are known, the
MCD model provides a measure of the innermost extent of the accretion
disk.  Assuming a distance of 2.5~kpc and $i=35^{\circ}$, and
$M_{BH}=10~M_{\odot}$, our fits indicate the inner disk may extend
very close to the marginally stable circular orbit: $R_{in}=8.7\pm
0.2~R_{g}~({\rm where}~ R_{g}=GM_{BH}/c^{2}$) via Model 3.  As
$R_{in}$ scales directly with the source distance in the MCD model, if
the distance to Cygnus X-1 is uncertain at the 25\% level the inner
disk extent should have approximately the same fractional uncertainty.
Errors in the inclination are less important, but non-negligible for
intermediate values.  Therefore, we regard the inner disk extent
measured via the MCD model to be consistent with the marginally stable
circular orbit for a Schwarzschild black hole ($6~R_{g}$) for a small
range of masses near 10~$M_{\odot}$.

Shimura and Takahara (1995) have proposed corrections to the MCD model
to account for Comptonization of the inner disk flux.  They suggest
that $kT_{in,corrected}=kT_{MCD}/f$ and $R_{in,corrected}=f^{2}\times
R_{in,MCD}$; $f=1.7$ is suggested as appropriate for Galactic black
holes.  If such a correction is valid, the inner disk may be somewhat
larger than the marginally stable orbit in the intermediate state.
Merloni, Fabian, \& Ross (2000) have noted that the MCD model may
systematically underestimate the inner disk extent and imply a
changing inner disk radius at low $\dot{m}$, but yield acceptable
measures at relatively high $\dot{m}$.  As we have observed Cygnus X-1
at a soft X-ray luminosity approximately twice as high as its
persistent luminosity, the MCD model may give acceptable estimates of
the inner disk extent.  In principle, the Fe~K$\alpha$ line can serve
as a check on the inner accretion disk extent.  The FWHM of the broad
Fe~K$\alpha$ line we discuss below is consistent with Keplerian
velocities expected if the inner accretion disk extends near to the
marginally stable circular orbit, suggesting $f\simeq 1.0$ may be more
appropriate in this state than $f=1.7$.

\subsection{The Fe~$K\alpha$ Line Region}
The complexity of the Fe~K$\alpha$ line region can be seen clearly
via two methods.  In Figure 3, the fit to the spectrum in the
Fe~K$\alpha$ line region with Model 3 is shown in detail.  In
Figure 4, we show the data/model ratio for a model which does not
include components to fit the Fe~K$\alpha$ line region, following
the procedure that Iwasawa et al. (1996) used to represent the
Fe~K$\alpha$ line in MCG~--6--30--15.

Relative to a model with no component to fit the narrow emission line,
the narrow Gaussian included in Model 1 is significant at the
$6.0\sigma$ level of confidence in the 0.65--10.0~keV band and above
$8.0\sigma$ in the 1.8--10.0~keV band.  The line measured via Model
1 is centered at $6.415\pm 0.007$~keV; this value is fixed in Models 3
and 4.  If this line were due mostly to Fe~I, this centroid energy
would represent a blue-shift of $560\pm 330$~km/s.  The FWHM width of
the line varies slightly depending on the underlying continuum model,
but is easily resolved with the HEG.  Via Model 3, we measure a FWHM
of $59^{+24}_{-14}$~eV, or $2800^{+1100}_{-660}$~km/s; a FWHM of
$80^{+28}_{-19}$~eV, or $3700^{+1300}_{-890}$~km/s is obtained via
Model 1.  The equivalent width of the line also depends slightly on
the underlying continuum: via Model 1 we obtain W$=22\pm 4$~eV; Model
3 gives W$=16^{+3}_{-2}$~eV.

The apparent blue-shift of the narrow line from Fe~I at 6.401~keV may
be explained in terms of a line mostly comprised of Fe~II, and
partially comprised of species below Fe~IX (Kaastra \& Mewe 1993).
Similarly, the measured FHWM of the line can be partially explained in
these terms.  Alternatively, one may want to argue that the line is
due mostly to Fe~I, and that the blue-shift and FWHM velocities are
physical.  However, the blue-shift is less than the terminal velocity
expected for a type-O stellar wind (1500~km/s; Castor, Abbot, \& Klein
1975); it is reasonable to assume that a neutral part of the wind must
be relatively far from the ionizing flux originating near the black
hole, and therefore close to terminal velocity.  Moreover, the blue
shift is far less than the jet velocity ($v/c > 0.6$) inferred from a
spectral model by Stirling et al. (2001) in the low/hard state of
Cygnus X-1.  It is more likely that the shift and FWHM are partially
explained by a line produced by a few moderately ionized species at a
point more internal to the system.

The measured strength of the narrow Fe K$_{\alpha}$ line is consistent
with \textit{ASCA} measurements of 10--30~eV by Ebisawa et al. (1996),
made during low/hard states.  Assuming solar Fe abundances, Ebisawa et
al. estimated that the maximum narrow line equivalent width expected
due to excitation of the companion star surface and wind is
$\leq11.1$~eV.  On this basis, Ebisawa et al. suggested that the
narrow line was most likely due to the irradiation of the outer
accretion disk.  This scenario is consistent with the reflection
geometry many authors have claimed in Cygnus X-1 (see, e.g.,
Gier\l{}inski et al. 1997, 1999).  Schulz et al. (2001) measured Fe to
be 25\% under-abundant in Cygnus X-1 ($A_{Fe}/A_{H} \simeq 3.3 \times
10^{-5}$; as per Morrison \& McCammon 1983 ); if this under-abundance
is intrinsic to the system, the maximum expected line equivalent width
due to excitation of the companion wind and surface is $\leq8.3$~eV.
We conclude that approximately half of the strength of the line we
observe must be produced via other means.  We conclude that the
irradiation of the cool outer accretion disk is likely to account for
the extra line strength.

Upper-limits on the strength of an Fe~K$\beta$ emission line at
7.06~keV (assuming the same FWHM measured for the K$\alpha$ line)
are not very constraining.  The broad emission line and smeared edge
both contribute at 7.06~keV, and these components may obscure a weak
K$\beta$ line.  The K$\beta$/K$\alpha$ line ratio is
consistent with the expected value of 0.13.

We regard the broad Gaussian plus smeared edge of Model 3, and
diskline plus smeared edge in Model 4, as approximations to a
self-consistent treatment of the Doppler and GR effects expected near
to a black hole, and to broadening effects expected if reflection is
important.  The smeared edge in Models 2, 3, and 4 has a fixed width
of 7.0~keV, as observed in other Galactic black holes (see, e.g.,
Sobczak et al. 1999).  We fix the smearing width because the energy
range of Chandra is not sufficient to constrain this parameter.  The
edge energy is fixed at 7.11~keV (for Fe~I) in Model 2, but is allowed
to vary in Models 3 and 4 (edge energies of $7.3\pm 0.1$~keV and
$7.2\pm 0.1$~keV are measured, respectively).  In Model 3, the maximum
optical depth is $\tau = 1.0\pm 0.2$.  The addition of the smeared
edge component in Model 2 is significant at the $5.2\sigma$ level of
confidence in the 0.65--10.0~keV band, and at $7.2\sigma$ in the
1.80--10.0~keV band.  

The addition of the diskline component Model 4 is only significant the
$3.1\sigma$ level of confidence in the 0.65--10.0~keV band, but is
significant at more than $8.0\sigma$ in the 1.80--10.0~keV band.  The
centroid energy measured is $5.85\pm 0.06$~keV.  The inclination was
moderately-well constrained: $i=40^{\circ}\pm 10^{\circ}$.  We note
that this measurement is broadly consistent with optical measurements
of the inclination ($i = 35^{\circ}$, Gies \& Bolton 1986, Herrero et
al. 1995).  Assuming an accretion disk emissivity profile with an
outer line excitation radius of $1000~R_{g}$, the inner disk extent
measured by the inner disk model is $R_{in}=7^{+6}_{-1}~R_{g}$.  This
value is consistent with the values we measured with the MCD model and
consistent with the marginally stable orbit around the black hole.
Note that this model does not assume a black hole mass, or a distance
to the source, which are necessary to derive inner disk radii in units
of $R_{g}$ with the MCD model.  Fits with the ``Laor'' model (Laor
1991) for a line produced at the inner accretion disk around a Kerr
black hole did not require an inner disk edge inside the marginally
stable circular orbit for a Schwarzschild black hole.  

Model 3 provides a better fit to the data, using a simple Gaussian to
model (see Figure 3) a broad Fe~K$\alpha$ emission line.  This
Gaussian component is significant at the $4.3\sigma$ level of
confidence in the 0.65--10.0~keV band, and at more than $8.0\sigma$ in
the 1.80--10.0~keV band.  The measured Gaussian centroid energy is
$5.82^{+0.06}_{-0.07}$~keV and the FWHM is very broad:
$1.9^{+0.5}_{-0.3}$~keV.  The line is relatively strong:
W$=140^{+70}_{-40}$~eV, but consistent with many previous measurements
of this feature.  As strong Doppler shifts and gravitational
red-shifts may be expected from the inner accretion disk, it is likely
that this line is produced via irradiation of the inner accretion
disk.

We have addressed the significance of the broad Fe~K$\alpha$ emission
line statistically, but an additional note is merited.  One would
expect that if an instrumental response error falsely creates a broad
line profile, that it should be seen in all observations of bright
sources made with \textit{Chandra}.  A review of the literature
reveals that this is not the case.  Moreover, \textit{Chandra}
observations of the Galactic black holes XTE~J1550 $-$564 at a flux of
0.6~Crab using our instrument mode (Miller et al. 2001d), and
GRS~1915$+$105 at a flux of 0.4~Crab using a standard instrument mode
(Lee et al. 2001), do \textit{not} reveal clear evidence for narrow or
broad Fe~K$\alpha$ emission lines.  These facts give us additional
confidence that the line we have observed is intrinsic to Cygnus~X-1,
and not due to the instrumental response.

The FWHM of the broad line implies Doppler shifts which are a
significant fraction of \textit{c} and consistent with Keplerian
velocities near the marginally stable orbit around a Schwarzschild
black hole.  As noted above, gravitational effects may also shape the
line profile.  However, other effects may be important.  As
Fe~K$\alpha$ line photons produced through disk reflection propagate
through the disk and/or an ionized disk skin, they may undergo Compton
scattering.  Here, we estimate the degree of line broadening due to
this process.  The broadening per scattering is given by:
$\frac{\Delta {\rm E}}{\rm E} = (\frac{2kT}{m_{e} c^2})^{0.5}$.
The inner accretion disk color temperature we have measured with the
MCD model is $<0.3$~keV.  RFY find that an thin ionized atmosphere
above the disk in Cygnus X-1 may have a temperature of approximately
$kT\sim 1.3$~keV.  These values give a broadening per scatter of 3\%
and 7\%, respectively.  For an Fe~K$\alpha$ line with a FWHM similar
to that of the narrow line we observe, approximately 50 scattering
events in a $kT\sim 1.3$~keV disk skin are required to reproduce the
full width of the broad line we observe.  For $\xi=10^{4}$ (slightly
above what we observe in Cygnus~X-1 in this state; see Section 1), RFY
find that Fe~XXV has a maximum ion fraction for $\tau_{Thomson}\sim
5$, suggesting as many as $\sim 25$ scattering events.  A hot coronal
volume (with $kT\sim 30$~keV, or higher) is not expected to contribute
to line broadening significantly as such volumes are likely to be
optically thin.  As the FWHM of the line we observe corresponds to
$\Delta {\rm E}/{\rm E} \sim 0.3$, for a range of disk temperatures
and ionized skin temperatures Compton scattering is likely to be a
relatively small but non-negligible broadening mechanism compared to
Doppler shifts and gravitational effects.

\section{Discussion}
A combination of an elevated source intensity, the resolution of the
\textit{Chandra} HETGS, and a $\sim$30~ks exposure have allowed us to
resolve the Fe~K$\alpha$ line region in the X-ray spectrum of
Cygnus X-1 for the first time.  We clearly detect a narrow
Fe~K$\alpha$ emission line, a broad Fe~K$\alpha$ emission line,
and a smeared Fe~K edge (see Figures 3 and 4).

These features can be explained in terms of an accretion disk
illuminated by a source of hard X-rays, with the broad line and edge
due to irradiation of the inner accretion disk, and the narrow line
due to irradiation of the outer accretion disk.  This scenario is
predicted by reflection models for AGNs and Galactic black holes (see,
e.g., George \& Fabian 1991; Ross, Fabian, \& Young 1999, Nayakshin \&
Dove 2001).  The broad components are consistent with previous
observations of Cygnus X-1 (see Ebisawa et al. 1996, Cui et al. 1998,
and Frontera et al. 2001, among others).  The narrow line is
consistent with upper-limits from \textit{ASCA} (Ebisawa et al. 1996).

The broad emission line is broader than that reported in most previous
observations of Cygnus X-1 and other Galactic black holes and may be
slightly red-shifted.  The broad line is not clearly double-peaked or
skewed like those observed in AGNs with \textit{ASCA} (for a review,
see Weaver, Gelbord, \& Yaqoob 2001; Yaqoob et al. 2002), or in an
\textit{XMM-Newton} observation of MCG~--6--30--15 by Wilms et
al. (2001).  The broad line has a centroid energy of
E$=5.82^{+0.06}_{-0.07}$~keV, which differs significantly from Fe~I at
6.40~keV.  The line profile we have observed may be regarded as
evidence for a line which is shifted and/or partially shaped by strong
gravitational effects.

Previous observations of Cygnus X-1 with the \textit{Chandra} HETGS
have not clearly revealed similar structure in the Fe~K$\alpha$
line region.  Schulz et al. (2001) observed this source at a similar
flux level for 15~ks (half of the exposure time we report on),
though with an observing mode less suited to this work and suffering
heavily from CCD pile-up, forcing the use of higher-order spectra with
lower sensitivity.  Cygnus X-1 was observed for 15~ks in the
low/hard state, at a flux approximately half of the intensity we
measured.  Preliminary results have been reported by Marshall et
al. (2001b).  The low/hard state observation does not reveal a narrow
emission line, but may show evidence for a weak, broad emission line
and smeared edge.  If the outer accretion disk is relatively more
flared in the intermediate state than the low/hard state, this could
explain why a narrow line is only detected in our observation.  The
lack of clearly detected lines in other \textit{Chandra} observations
of Cygnus X-1 may indicate that the line is variable.  However, it is
possible that long exposures in the future with well-suited instrument
modes will reveal structure in the Fe~K$\alpha$ line region across
a range of source intensities.

Previous observations of transient Galactic black holes have also
revealed evidence for broad Fe~K$\alpha$ emission lines, though few
are as clear as the profiles observed in Cygnus X-1.  Among the
sources recently observed, broad lines have been been detected in XTE
J1550$-$564 (Sobczak et al. 2000), GRO~J1655$-$40 (Sobczak et
al. 2000, Ba\l{}uci\'nska-Church \& Church 2000), GX~339$-$4 (Nowak,
Wilms, \& Dove 2002), XTE J1748$-$288 (Miller et al. 2001e), V4641~Sgr
(in't Zand et al. 2000), and XTE~J2012$+$381 (Campana et al. 2001).
These lines have been observed in the ``very high'' state and in
intermediate states.  In both of these states, the inner accretion
disk temperature in these transient systems is relatively high:
$kT=0.7-1.2$~keV is common.  Moreover, the inclinations of
GRO~J1655$-$40 (Greene, Bailyn, \& Orosz 2001) and XTE~J1550$-$564
(Orosz et al. 2001) are $i=70^{\circ} \pm 2^{\circ}$ and $i=72^{\circ}
\pm 5^{\circ}$, respectively.  The relatively cool inner disk we have
observed in the intermediate state of Cygnus X-1 ($kT=0.236\pm
0.002$~keV) makes continuum flux from the disk less important in the
Fe~K$\alpha$ line region than in transient systems.  The low
inclination ($i\simeq 35^{\circ}$) of Cygnus X-1 may also reveal
irradiation of the inner disk more clearly than transient black holes
with high inclinations.  

The Fe~K$\alpha$ line provides an important diagnostic of the
innermost extent of the accretion disk.  The breadth of the line and
edge we have observed --- if produced at the inner accretion disk ---
require a disk which extends close to the marginally stable circular
orbit.  The line profile is consistent with the values of the inner
disk extent we have measured with the MCD model.  This finding
supports the ionized disk model of Young et al. (2001) for X-ray
states in Cygnus X-1 (for another discussion of ionized transition
layers, also see \.Zycki, Done, \& Smith 2001).  The inner disk extent
in the intermediate state may be particularly incisive in evaluating
the ADAF model.  Esin et al. (1998) found that the disk may extend to
the marginally stable orbit in the high/soft state of Cygnus X-1, but
may have recessed to $200~R_{g}$ in the low/hard state.  That the
inner disk extent is consistent with the marginally stable orbit in
this intermediate state suggests the inner disk may not recede
smoothly as a function of the mass accretion rate in Cygnus X-1.  In
commenting on the ionized disk and ADAF models, it must be noted that
within the larger set of states and behaviors observed in Cygnus X-1,
this state may prove to be peculiar.  Several future observations of
Cygnus X-1 at high resolution are required to further evaluate these
models.

\section{Conclusions}
The main results of this paper may be summarized as follows:

$\bullet$ We have resolved the Fe~K$\alpha$ line region into a narrow
line consistent with the excitation of low ion species, and a very
broad emission line and edge combination.  The lines are likely
produced via the irradiation of the accretion disk, with the broad
line produced at the inner accretion disk and the narrow line excited
at the outer accretion disk.  This is consistent with models for X-ray
reflection in Galactic black holes and AGNs.  The broad line shape may
be determined predominantly by a combination of Doppler shifts, the
gravitational field of the black hole, and also by Compton scattering
in the accretion disk and/or an ionized disk skin as part of the
reflection process.

$\bullet$ Based on the accretion disk temperature, the photon
power-law index, and the X-ray flux observed, we conclude that we
observed Cygnus X-1 in an intermediate state.

$\bullet$ In this intermediate state, the MCD model suggests that the
inner accretion disk extends close to the marginally stable orbit.
This finding is supported by the broad Fe~K$\alpha$ line and edge
profiles.  Thus, an inner ADAF is not required to describe this state
of Cygnus X-1.

$\bullet$ When well-suited observational modes and long exposures are
used to observe bright galactic sources, the \textit{Chandra} HETGS is
capable of resolving composite lines into components.  This result
holds great promise for understanding the accretion flow geometry in
Galactic black holes and neutron stars via Fe~K$\alpha$ line
diagnostics.

\section{Acknowledgments}
We wish to thank \textit{Chandra} Director Harvey Tananbaum, and the
\textit{Chandra} staff for executing this observation and their help
in processing the data.  We thank Michael Nowak for useful
discussions.  RW was supported by NASA through Chandra fellowship
grants PF9-10010, which is operated by the Smithsonian Astrophysical
Observatory for NASA under contract NAS8--39073.  HLM and NSS are
supported under SAO contract SAO~SV1-61010.  WHGL gratefully
acknowledges support from NASA.  This research has made use of the
data and resources obtained through the HEASARC on-line service,
provided by NASA-GSFC.

\clearpage


\begin{table}[t]
\caption{Models for the 0.65-10.0 keV Spectrum of Cygnus X-1}
\begin{scriptsize}
\begin{center}
\begin{tabular}{llllll}
\multicolumn{2}{l}{~} & Model 1 & Model 2 & Model 3 & Model 4\\
\tableline
\multicolumn{2}{l}{(MCD$^{a}$)} & ~ & ~ & ~ & ~\\
\multicolumn{2}{l}{$kT$ (keV)} & 0.240(3) & 0.241(2) & 0.236(2) & 0.239(2) \\
\multicolumn{2}{l}{Norm. ($10^{5}$)} & 2.04(8) & 1.99(7) & 2.14(9) & 2.07(7) \\
\multicolumn{2}{l}{R$_{in}$ (R$_{g}$)} & 8.5(2) & 8.4(2) & 8.7(2) & 8.5(2) \\
\multicolumn{2}{l}{Flux ($10^{-8}$ cgs)} & 0.36(1) & 0.36(1) & 0.35(1) & 0.36(1) \\
\tableline
\multicolumn{2}{l}{(Power-law)} & ~ & ~ & ~ & ~\\
\multicolumn{2}{l}{$\Gamma$} & 1.789(9) & 1.78(1) & 1.84(2) & 1.80(1) \\
\multicolumn{2}{l}{Norm.} & 2.009(9) & 1.98(2) & 2.09(2) & 2.03(2) \\
\multicolumn{2}{l}{Flux ($10^{-8}$ cgs)} & 1.040(5) & 1.014(8) & 1.04(1) & 1.03(1) \\
\tableline
\multicolumn{2}{l}{L$_{X}^{a}$~($10^{37}$~cgs)} & 1.05(2) & 1.03(2) &
1.03(2) & 1.04(2) \\
\tableline
\multicolumn{2}{l}{(Narrow Line)} & ~ & ~ & ~ & ~\\
\multicolumn{2}{l}{E~(keV)} & 6.415(7) & $6.415^{+0.007}_{-0.003}$ & 6.415 & 6.415 \\
\multicolumn{2}{l}{$\sigma$~(eV)} & $34^{+12}_{-8}$ & $36^{+6}_{-11}$ & $25^{+10}_{-6}$ & 30(6) \\
\multicolumn{2}{l}{W~(eV)} & 22(4) & 20(4) & $16^{+3}_{-2}$ & 22(3) \\
\multicolumn{2}{l}{Norm. ($10^{-3}$)} & 1.6(3) & $1.6^{+0.2}_{-0.4}$ & 1.2(3) & 1.6(2) \\
\multicolumn{2}{l}{Flux ($10^{-11}$~cgs)} & 1.7(3) &
$1.7^{+0.2}_{-0.4}$ & 1.2(3)  & 1.7(2) \\
\tableline
\multicolumn{2}{l}{(Broad Line)} & -- & -- & (Gaussian) & (Diskline$^{c}$)\\
\multicolumn{2}{l}{E~(keV)} & -- & -- & $5.82^{+0.06}_{-0.07}$ & 5.85(6) \\
\multicolumn{2}{l}{$\sigma$~(keV)} & -- & -- & $0.8^{+0.2}_{-0.1}$ & -- \\
\multicolumn{2}{l}{W~(eV)} & -- & -- & $140^{+70}_{-40}$ & $60^{+12}_{-6}$ \\
\multicolumn{2}{l}{Norm. ($10^{-3}$ cgs)} & -- & -- & $12^{+6}_{-3}$ & $4.5^{+0.9}_{-0.4}$ \\
\multicolumn{2}{l}{Flux ($10^{-11}$ cgs)} & -- & -- & $12^{+6}_{-3}$ & $4.0^{+0.8}_ {0.4}$\\
\multicolumn{2}{l}{R$_{in}$ (R$_{g}$)} & -- & -- & -- & $7^{+6}_{-1}$ \\
\multicolumn{2}{l}{$i$ (degrees)} & -- & -- & -- & 40(10) \\
\tableline
\multicolumn{2}{l}{(Smeared Edge)} & ~ &  & ~ & ~\\
\multicolumn{2}{l}{E~(keV)} & -- & 7.11 & 7.3(2) & 7.2(1) \\
\multicolumn{2}{l}{$\tau$} & -- & 1.5(3) & 1.0(2) & 1.2(2) \\
\tableline
\multicolumn{2}{l}{(1.80--10.0~keV)} & ~ &  & ~ & ~\\
\multicolumn{2}{l}{$\chi^{2}$} & 1176.94 & 1130.69 & 1104.20 & 1114.45 \\
\multicolumn{2}{l}{d.o.f.} & 1293 & 1292 & 1289 & 1287 \\
\multicolumn{2}{l}{P$_{F-statistic}^{b}$} & $8.5\times 10^{-21}$ & $6.2\times 10^{-13}$ & $2.6\times 10^{-19}$ & $5.0\times 10^{-18}$ \\
\multicolumn{2}{l}{~} & ($>8.0\sigma$) & ($7.2\sigma$) & ($>8.0\sigma$) & ($>8.0\sigma$) \\
\tableline
\multicolumn{2}{l}{(0.65-10.0~keV)} & ~ &  & ~ & ~\\
\multicolumn{2}{l}{$\chi^{2}$} & 6576.86 & 6527.52 & 6482.62 & 6493.93 \\
\multicolumn{2}{l}{d.o.f.} & 3614 & 3613 & 3610 & 3608 \\
\multicolumn{2}{l}{P$_{F-statistic}^{b}$} & $2.0\times 10^{-9}$ &
$1.8\times 10^{-7}$ & $1.6\times 10^{-5}$ & $2.3\times 10^{-3}$ \\
\multicolumn{2}{l}{~} & ($6.0\sigma$) & ($5.2\sigma$) & ($4.3\sigma$) & ($3.1\sigma$) \\
\tableline
\end{tabular}
\vspace*{\baselineskip}~\\ \end{center} \tablecomments{Errors on the
MCD and power-law components are 90\% confidence errors, and errors on
line parameters are $1\sigma$ errors.  Where errors are symmetric,
they are indicated in parentheses.  Single-digit errors reflect the
error in the last significant digit.  Where errors are not quoted, the
parameter was fixed at the value indicated.  The ISM absorption model
of Schulz et al. (2001) was used in fitting the spectra.  $^{a}$ We
assume $i=35^{\circ}$ and $d=2.5$~kpc (see text for references).
$^{b}$ P is the F-statistic probability that the improvement in the
$\chi^{2}$ fitting statistic is due to random fluctuations.  For Model
1, P is quoted for the addition of the Gaussian model for the narrow
Fe~K$\alpha$ line to the same model with no lines.  For Model 2, P
relates to the addition of the smeared Fe~K edge model ``smedge''
versus Model 1.  For Models 3 and 4, P is quoted for the addition of
Gaussian and ``diskline'' models (respectively) for the broad
Fe~K$\alpha$ line component, relative to Model 2.  P was calculated
using the ``ftest'' task within XSPEC version 11.1.0 (Arnaud 1996).
Underneath each P value, the significance of the feature is indicated
in parentheses.  The apparently poor fits in the 0.65--10.0~keV band
($\chi^{2}/\nu \sim 1.80$) are due to complex line spectrum below 2~keV.
We also quote the results of fitting these models on the 1.8--10.0~keV
range, which avoids much of the absorption but allows the instrumental
feature at 2.0~keV to be constrained.  $^{c}$ An accretion disk
emissivity profile was assumed for the diskline model.  }
\vspace{-1.0\baselineskip}
\end{scriptsize}
\end{table}

\clearpage

\begin{figure}
\figurenum{1}
\label{fig:asm}
\centerline{~\psfig{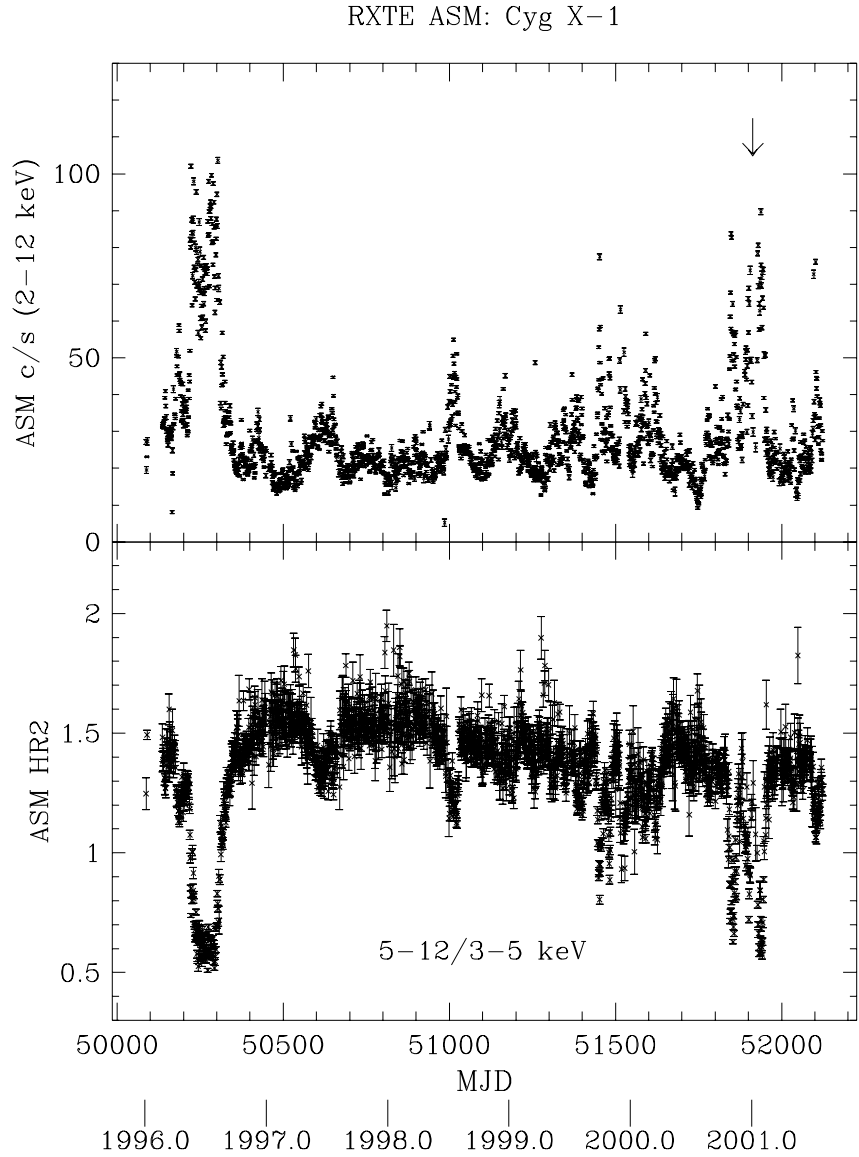}~}
\caption{\footnotesize The \textit{RXTE}/ASM lightcurve and
(5-12~keV)/(3-5~keV) hardness ratio for Cygnus~X-1.  The curves shown
here span a time interval from the start of the mission until shortly
after our \textit{Chandra} observation on 2001 January 4 (indicated by
the arrow in the lightcurve).  On this day, Cygnus X-1 was observed at
34~c/s with the ASM.  The source was the midst of the softest,
highest-intensity period observed until that point since the 1996
high/soft state.}
\end{figure}

\begin{figure}
\figurenum{2}
\label{fig:Broadband}
\centerline{~\psfig{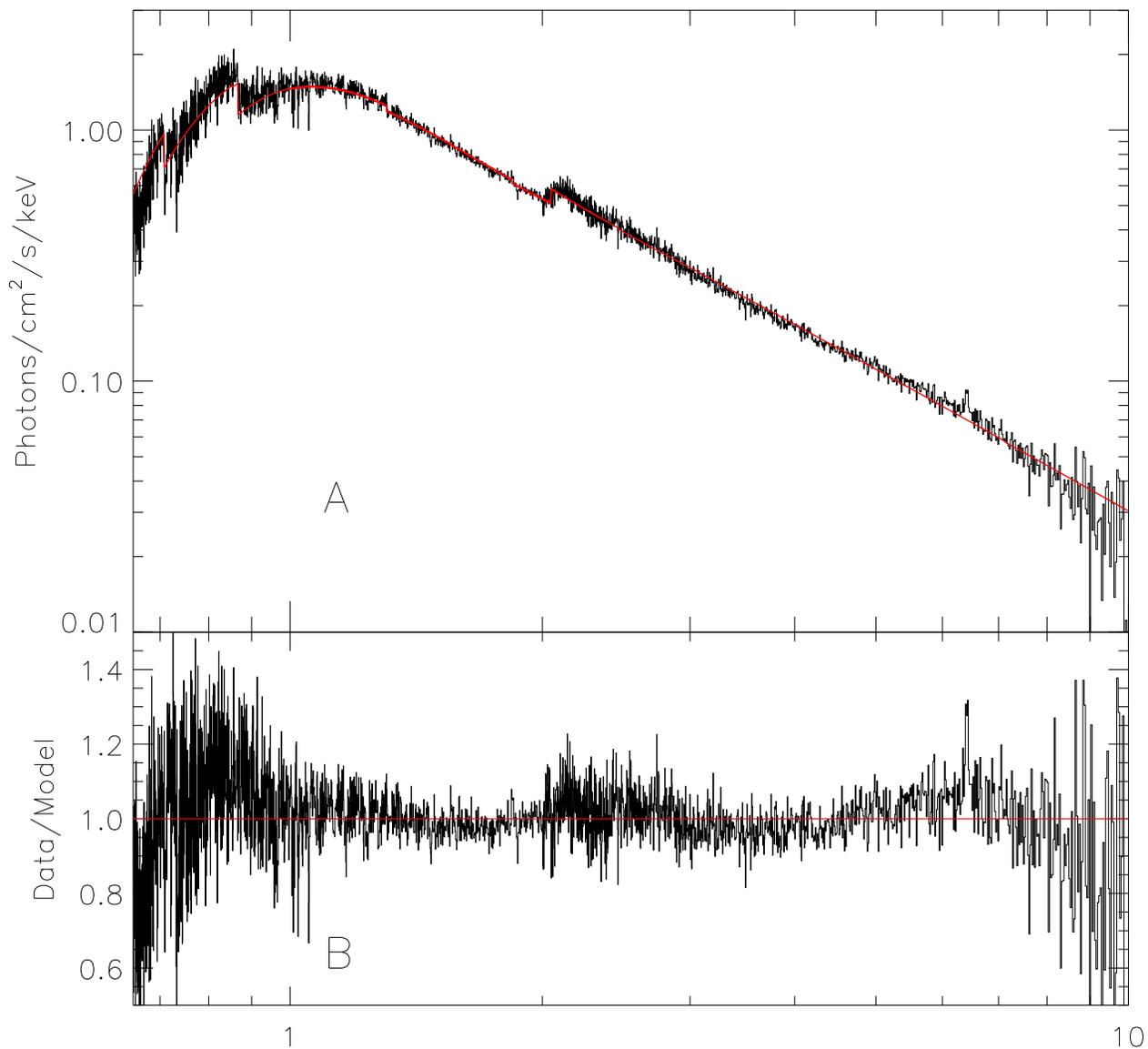}~}
\caption{\footnotesize In panel A, the 0.65-10.0~keV spectrum is fit
with a multicolor disk blackbody plus power-law continuum model (Model
1 in Table 1).  The data/model ratio for this model is shown in panel
B.  We have suppressed a narrow instrumental feature at 1.85~keV.  The
notch at 2.05~keV is also an instrumental artifact.  To portray the
spectrum with greatest clarity, we have plotted the spectrum and
ratio without errors.  However, all errors are small compared to the
deviations except below 0.7~keV and above 8.5~keV.}
\end{figure}

\begin{figure}
\figurenum{3}
\label{fig:Fe}
\centerline{~\psfig{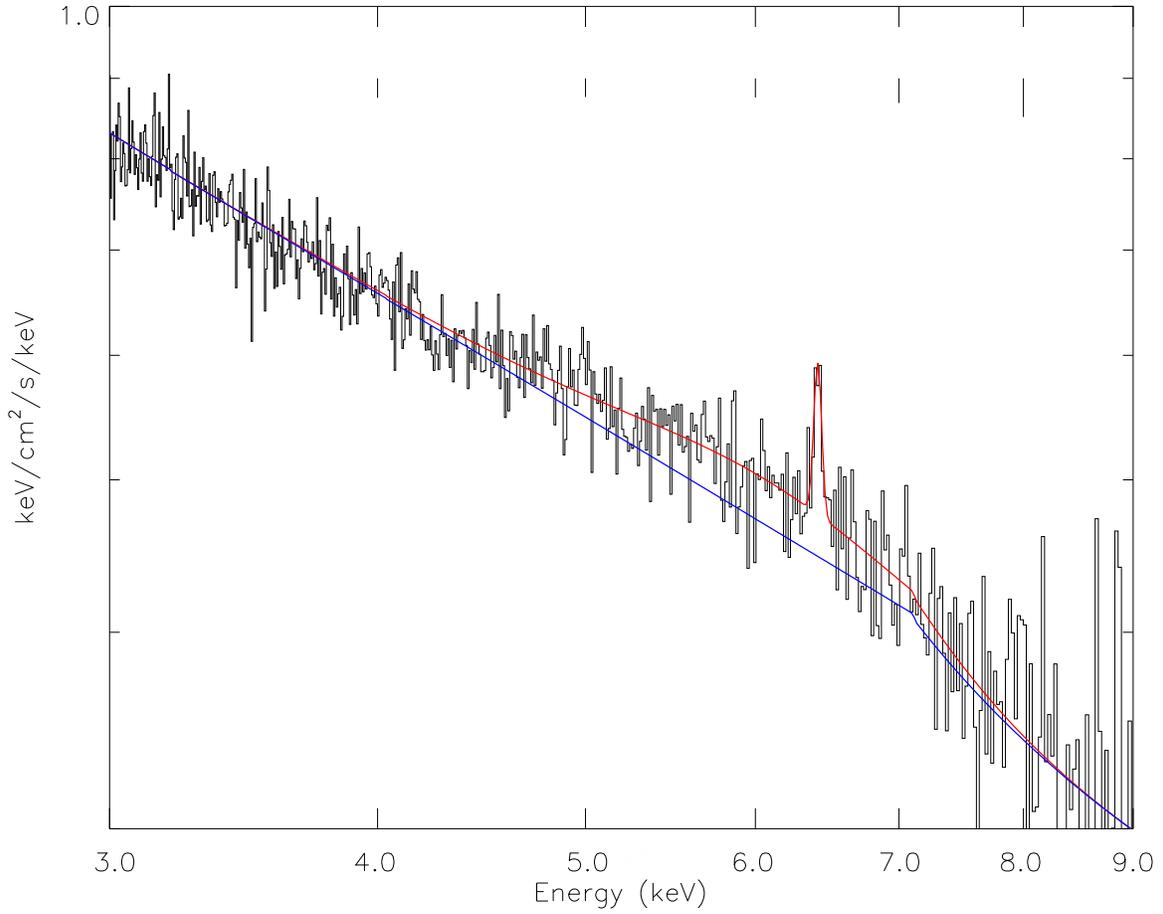}~}
\caption{\footnotesize The HEG spectrum in the Fe~K$\alpha$ line
region at full resolution.  The fit in red corresponds to Model 3 in
Table 1 (the continuum model is that of a multicolor disk blackbody
plus a power-law).  The fit includes broad and narrow Gaussian line
components and a smeared Fe~K edge.  The continuum
\textit{not} including the lines is shown by the fit in blue.
Representative error bars are indicated along the top of the plot.
The resolution of the HEG reveals that the Fe~K$\alpha$ line in
Cygnus X-1 is a composite of narrow and broad lines.}
\end{figure}

\nopagebreak

\begin{figure}
\figurenum{4}
\label{fig:ratio}
\centerline{~\psfig{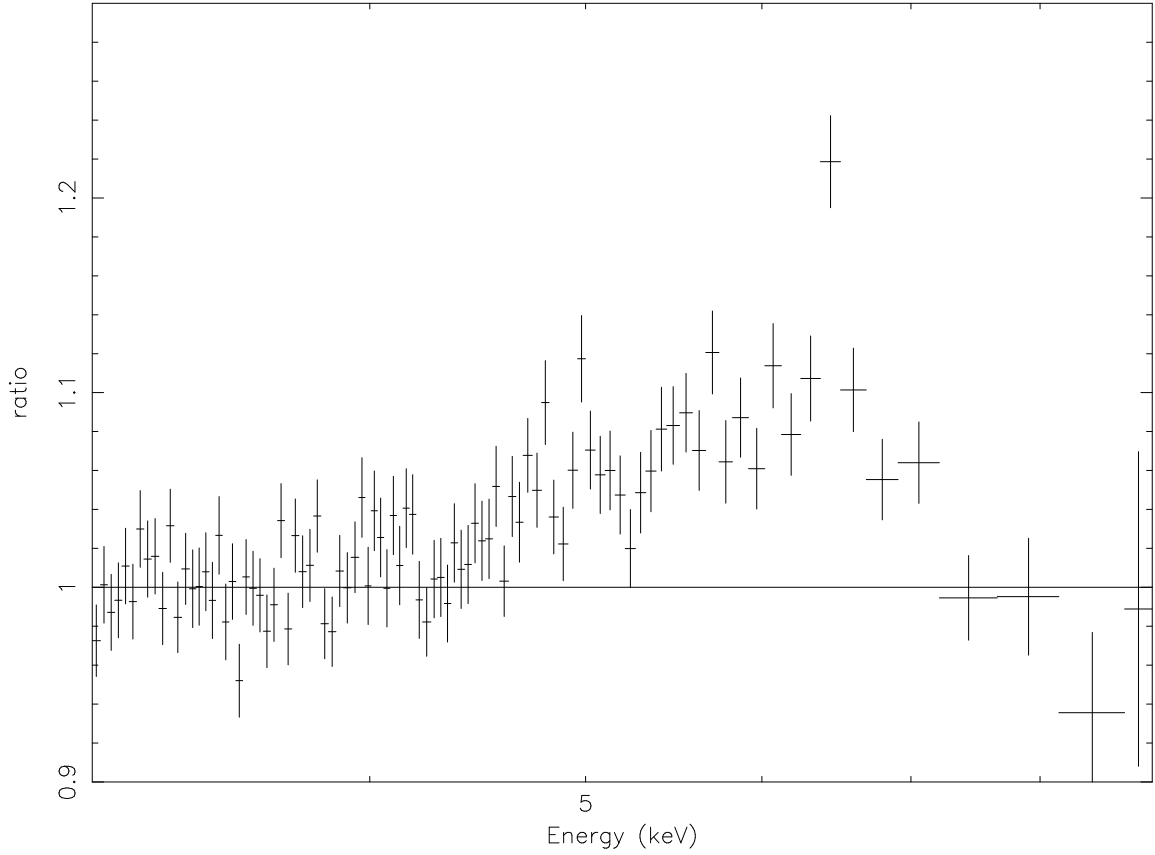}~}
\caption{\footnotesize The data/model ratio of the Fe~K$\alpha$
line region, rebinned by a factor of 50 for visual clarity.  The
spectrum was fit with a simple multicolor disk blackbody plus
power-law model (see Table 1), but ignoring the 4.0--7.2~keV region
(following Iwasawa et al. 1996 for the Seyfert galaxy MCG
--6--30--15).  The data/model ratio above clearly displays an iron
line profile which is very similar to profiles seen in active galactic
nuclei.}
\end{figure}

\end{document}